\documentclass[12pt,preprint]{article}%
\usepackage{amssymb}
\usepackage{amsmath}
\usepackage{graphics}
\usepackage{epsfig}
\usepackage{amsfonts}
\usepackage{graphicx}%
\renewcommand{\vec}[1]{{\bf #1}}
\newcommand{\eqb}{\begin{equation}}
\newcommand{\eqe}{\end{equation}}
\newcommand{\dmb}{\begin{displaymath}}
\newcommand{\dme}{\end{displaymath}}
\newcommand{\pd}{\partial}

\newcommand{\eab}{\begin{eqnarray}}
\newcommand{\eae}{\end{eqnarray}}

\newcommand{\e}{\mbox{e}}
\newcommand{\be}{\begin{equation}}
\newcommand{\ee}{\end{equation}}

\begin{document}
\begin{titlepage}
\begin{flushright}
\end{flushright}
\vspace{0.6cm}
\begin{center}
\Large{Electromagnetic waves and photons}  
\vspace{0.5cm}\\ 
\small{Ralf Hofmann}
\end{center}
\begin{center}
{\em Institut f\"ur Theoretische Physik\\
Universit\"at Heidelberg\\
Philosophenweg 16\\
69120 Heidelberg, Germany}
\end{center}
\vspace{1.5cm}

\begin{abstract}

We explore how the thermal ground states of two mixing and 
pure SU(2) Yang-Mills theories, SU(2)$_{\tiny\mbox{CMB}}$ 
of scale $\Lambda_{\tiny\mbox{CMB}}\sim 10^{-4}\,$eV and 
SU(2)$_{e}$ of scale $\Lambda_{e}\sim 5\times 10^5\,$eV, associate either 
wave or particle aspects to electromagnetic disturbances during 
thermalisation towards the photon gas of a blackbody, in realising the photoelectric effect, 
and through the frequency dependence of the monochromatic, nonthermal beam structure in Thomson/Compton scattering.  

\end{abstract}
\vspace{0.5cm}
$\mbox{}^*$email: r.hofmann@thphys.uni-heidelberg.de

\end{titlepage}

In deriving the 
field equations of classical electrodynamics Maxwell assumed that 
electromagnetic disturbances are those of a medium -- the luminiferous aether -- required for them to 
propagate in analogy to distortions of the stationary flow of a fluid in hydrodynamics \cite{MaxwellTreatise}. 
If existent then such a medium needs to be of a peculiar nature, however, since it fails to 
associate with a preferred rest frame, a feat first demonstrated by Michelson and Morley 
\cite{MichelsonMorely} and re-confirmed many times ever since: the speed of light $c$ 
is a constant of nature and as such does not depend on the observer's 
state of motion relative to a source. Consequences of this experimental fact, expressed by the group of 
Lorentz transformations linking inertial frames,  
are laid out by Special Relativity \cite{EinsteinSR} and have 
been vindicated by countless 
experiments. In classical electrodynamics, constancy of the phase velocity $c$
of electromagnetic waves is implied by the constancy of $\epsilon_0$ and $\mu_0$ -- the 
electric permittivity and magnetic permability of free space. On the other hand, 
experience associates a particle-like or quantum nature 
to the carrier of the electromagnetic force -- the photon 
\cite{Planck1900,Einstein1905,Einstein1909} -- whose energy $E=2\pi\hbar\nu$ and momentum modulus $p=\frac{2\pi}{c}\nu$ are independent of 
the intensity of the monochromatic electromagnetic wave it associates with, but proportional to 
frequency $\nu$. Here $\hbar$ denotes   
the (reduced) fundamental quantum of action. From now  
on we use natural units $c=\hbar=k_B=\epsilon_0=\mu_0=1$, 
$k_B$ indicating Boltzmann's constant.      

The purpose of this note is to propose a framework, 
based on the extension of the gauge group of 
electromagnetism U(1) to a product of mixing SU(2) groups belonging to pure 
Yang-Mills theories. As we will argue, such a setting promises to reconcile 
the seemingly paradoxical wave-particle aspects of 
electromagnetic disturbances in terms of a nontrivial vacuum structure. In addressing the latter, the key 
instrument is the thermal ground state of SU(2) Yang-Mills theory. This concept 
is rooted in topologically nontrivial 
field configurations (trivial-holonomy Harrington-Shepard calorons and anticalorons \cite{HS1977} of topological charge-modulus unity) 
of period $\beta=1/T$ in the Euclidean 
time coordinate $\tau$, $T$ being a temperature parameter, and in a spatial coarse graining 
\cite{HerbstHofmann2004}. It is important to note that the derivation of the thermal ground state does not require the consideration of 
thermal excitations. Therefore, it should be generalisable to 
the description of isolated excitation modes of the effective gauge field such that parameter $T$ 
enjoys a nonthermal interpretation.       

Effectively, that is, after spatial coarse graining, the thermal ground state is 
characterised by an inert, adjoint scalar field $\phi$ and an effective, pure-gauge confi\-guration $a_\mu^{\tiny\mbox{bg}}$, 
the latter solving the Yang-Mills equations subject to a source 
term provided by the former \cite{Hofmann2005}. 
While field $\phi$ represents spatially densely packed caloron/anti\-caloron 
centers -- their dependence on Euclidean time 
$\tau$ coarse grained into a mere choice of gauge at the spatial scale $|\phi|^{-1}$ \cite{HerbstHofmann2004} --  
the effective gauge field $a_\mu^{\tiny\mbox{bg}}$ represents the collective effect 
of caloron/anti\-caloron overlap, accompanied by transient holonomy changes \cite{Nahm,VanBaal,LeeLu,Diakonov}, as facilitated by 
their static peripheries. Caloron/anticaloron peripheries set in at spatial 
scale $s=\pi\frac{|\phi|^{-2}}{\beta}$ \cite{Krasowski,GPY1982,GrandouHofmann2015} where 
\eqb
\label{sofT}
s(\lambda)=\frac12\lambda^2 \Lambda^{-1}\,,\ \ \ (\lambda\equiv 2\pi T/\Lambda)\,,
\eqe 
and $\Lambda$ denotes the Yang-Mills scale of the SU(2) theory. This introduces a 
finite energy density $\rho^{\tiny\mbox{gs}}=4\pi\Lambda^3 T$ to the 
thermal ground state \cite{Hofmann2005} which, due to the (anti)caloron's (anti)selfduality,  
would vanish in isolation \cite{SchaferShuryak}. 

For a given caloron, 
the peripheral field strength is that of a 
selfdual, static dipole\footnote{The case of a small caloron scale parameter $\rho=|\phi|^{-1}\ll\beta$ was also discussed in \cite{GPY1982}. 
Here the static and self\-dual dipole emerges for spatial distances $r\gg\beta$ and for 
$\lambda\ll (2\pi)^{2/3}$. Such a situation is, however, inconsistent with the deriva\-tion of the thermal ground state 
\cite{HerbstHofmann2004,Hofmann2005}.} \cite{GPY1982}. On the other hand, field $\phi$ breaks the 
SU(2) gauge symmetry of the underlying, classical Euclidean Yang-Mills action down to U(1) 
which means that only one of the three directions of the SU(2) algebra su(2) is massless, 
the (large) mass of the other two directions being fixed by (low-temperature) ambient thermodynamics in a large bulk volume 
\cite{Hofmann2005,lineTemp,NofNature2013}. 
A natural question to ask is under what conditions the associated electric and magnetic dipole 
densities of the thermal ground state can be regarded a medium induced by, at the same time 
supporting, wave-like propagation of the mass\-less mode. By promoting 
$a_\mu^{\tiny\mbox{bg}}$ (in unitary gauge: $\phi=2|\phi|t^3$, $|\phi|=\sqrt{\frac{\Lambda^3}{2\pi T}}$, 
generators $t^a$ $(a=1,2,3)$ normalised as tr\,$t^at^b=\frac12\,\delta^{ab}$) to a monochromatic 
electro\-magnetic wave $a^{a=3}_\mu$, associated with the massless su(2) direction\footnote{In SU(2)$_{\tiny\mbox{CMB}}$ massive modes $a^{a=1,2}_\mu$ interact with the massless mode $a^{a=3}_\mu$ by tiny radiative corrections only at 
a thermodynamical  temperature bounded from below by that of the present CMB \cite{Hofmann2005} and 
thus can be ignored in the effective Yang-Mills equation $D^\mu G_{\mu\nu}=2ie[\phi,D_\nu\phi]$ 
when discussing the propagation of $a^{a=3}_\mu$. This equations thus reduces to the vacuum Maxwell 
equation $\pd^\mu F^3_{\mu\nu}=0$ with $F_{\mu\nu}=\pd_\mu a^{a=3}_\nu-\pd_\nu a^{a=3}_\mu$ which, indeed, 
is solved by a plane wave. The latter is subject to undetermined normalisation, frequency, and phase. Note that, 
as is the case for $a_\mu^{\tiny\mbox{bg}}$, the Euclidean, time averaged energy density 
tr$\frac12(\vec{E}_e^2-\vec{B}_e^2)$ of $a^{a=3}_\mu$ vanishes 
such that solely the potential tr$V(\phi)=\rho^{\tiny\mbox{gs}}$ in the effective action determines the mean 
energy density of such a plane wave \cite{Hofmann2005}.}, and 
by identifying its mean Minkowskian energy density 
with $\rho^{\tiny\mbox{gs}}$, one arrives at \cite{GrandouHofmann2015} 
\eqb
\label{meanET}
|\vec{E}_e|=\Lambda^2\sqrt{2\lambda}\,,
\eqe
where $|\vec{E}_e|$ represents the mean electric field-strength modulus of $a^{a=3}_\mu$. 
Thus parameter $T=\frac{\lambda\Lambda}{2\pi}$ is set by the wave's intensity. Using this relation and exploiting that the dipole density is given by the ratio of dipole moment per (anti)caloron to the spatial coarse-graining volume, it was shown in 
\cite{GrandouHofmann2015} that the electric permittivity $\epsilon_0$ 
and the magnetic susceptibility $\mu_0$ of the thermal ground 
state are independent of $T$. 

Eq.\,(\ref{meanET}) together with the condition that wavelength $l$ must not resolve the 
interior of a caloron/anticaloron, $l\gg s(\lambda)$, imply the following $T$ independent statement \cite{GrandouHofmann2015}
\eqb
\label{uncert}
|\vec{E}_e|^4 l^{-1}=|\vec{E}_e|^4\nu\ll 8\Lambda^9\,,
\eqe
$\Lambda$ thus determines the maximum of 
intensity at a given frequency $\nu$ and vice versa commensurate with wave-like propagation. 
Although derived from the two $T$ dependent relations $l\gg s(\lambda)$, see (\ref{sofT}), 
and (\ref{meanET}) condition (\ref{uncert}) 
should be regarded universal. That is, in a 
nonthermal situation, $\nu$ is not required to satisfy any 
additional constraint as implied by the 
existence of a critical thermodynamical temperature $\lambda_c=13.87$ 
for the deconfining-preconfining phase transition \cite{Hofmann2005}.  

For SU(2)$_{\tiny\mbox{CMB}}$, the Yang-Mills 
factor proposed to underly all experimentally investigated 
thermal photon gases including the Cosmic Microwave Background (CMB) \cite{Hofmann2005}, one has $\Lambda_{\tiny\mbox{CMB}}\sim 10^{-4}\,$eV 
\cite{lineTemp}. According to (\ref{uncert}) the energy density 
$|\vec{E}_e|^2$ is bounded by $|\vec{E}_e|^2\ll \sqrt{8\,\frac{\Lambda^9_{\tiny\mbox{CMB}}}{\nu}}$. 
For $\nu=10^6\,$Hz (radio frequency) one obtains $|\vec{E}_e|^2\ll 2.3\times 10^{-21}\,$J\,cm$^{-3}$. For a comparison, 
the energy density of the CMB at this frequency, measured with a spectral band width of 
$\Delta\nu=10^4\,$Hz, is $8\pi T\nu^2\Delta\nu=3.51\times 10^{-37}\,$J\,cm$^{-3}$. 
Thus, such a radio wave could represent a signal discernible from 
the thermal noise of the CMB. Higher-frequency monochromatic waves 
are bounded by energy densities reduced by a 
factor $1/\sqrt{\frac{\nu}{10^6\,\tiny\mbox{Hz}}}$, and 
it is clear that condition (\ref{uncert}) is violated for a wealth of phenomena, attributed to 
the propagation of electromagnetic waves, when setting $\Lambda=\Lambda_{\tiny\mbox{CMB}}$. 

The way out is to postulate the existence of additional SU(2) factors 
\cite{Hofmann2005,GiacosaHofmann2007} with Yang-Mills scales hierarchically 
larger than $\Lambda_{\tiny\mbox{CMB}}$ which, thermodynamically seen, 
are in confining phases under ambient conditions such that massive modes do not propagate. 
One could consider $\Lambda=\Lambda_e\sim 
5\times 10^5\,$eV$\sim m_e$, $m_e$ denoting the electron mass. Then (\ref{uncert}) no longer is in 
contradiction with experience: propagation of high-intensity and 
high-frequency massless waves is accomodated by the large 
value of the Yang-Mills scale $\Lambda_e$. 

The process of thermalisation in SU(2)$_{\tiny\mbox{CMB}}$ towards blackbody radiation, 
which is surrounded by a cavity wall providing emitting and absorbing 
electrons, would then proceed as follows. At a 
thermodynamical wall temperature $T$ with $\Lambda_e>T\gg T_{\tiny\mbox{CMB}}=2.725\,$K$=\frac{13.87}{2\pi}\Lambda_{\tiny\mbox{CMB}}$ \cite{lineTemp} 
radiation emitted by the wall electrons satisfies (\ref{uncert}) with $\Lambda=\Lambda_e$. This radiation represents 
classical waves in SU(2)$_e$. A priori, their spectral energy density thus is given by the Rayleigh-Jeans 
law
\eqb\label{RJlaw}
u_{\tiny\mbox{RJ}}=8\pi T\nu^2=\frac{2}{\pi}\,T^3 x^2\,, \ \ (x\equiv 2\pi\nu/T)\,,
\eqe
which expresses an obvious and well-known ultraviolet catastrophe. The latter, however, does not take place if 
classical SU(2)$_e$ waves excite {\sl photons} from the thermal 
ground state of SU(2)$_{\tiny\mbox{CMB}}$. Namely, setting $\Lambda=\Lambda_{\tiny\mbox{CMB}}$ in a thermal situation, 
the condition that wavelength $l$ must be 
larger than $s$ (see Eq.\,(\ref{sofT})) for wave-like propagation amounts to 
\eqb
\label{violcmb}
l=\frac{2\pi}{xT}\gg s=\frac{2\pi^2 T^2}{\Lambda^3_{\tiny\mbox{CMB}}}\ \Leftrightarrow \ 
\ x\ll\frac{1}{\pi} \left(\frac{\Lambda_{\tiny\mbox{CMB}}}{T}\right)^3\,.
\eqe
Hence, condition (\ref{violcmb}) is violated at extremely small frequencies already, that is, for 
$x>\frac{1}{\pi} \left(\frac{\Lambda_{\tiny\mbox{CMB}}}{T}\right)^3$. 
For such frequencies the quantum of action, localised in thus probed caloron/anticaloron 
centers \cite{Krasowski,GrandouHofmann2015}, participates in the thermodynamics of fluctuations by 
indeterministic materialisations of quanta  
of energy and momentum $2\pi\nu$. In assuming that their numbers are suppressed by associated 
Boltzmann weights the Bose-Einstein distribution function $n_B(x)=1/(\e^x-1)$ is implied, corresponding to  
a spectral energy density  
\eqb\label{Planck}
u_{\tiny\mbox{Planck}}=\frac{2}{\pi}\,T^3 \frac{x^3}{\e^x-1}\,. 
\eqe 
Function $u_{\tiny\mbox{Planck}}$ peaks at $x=2.82$, is normalisable to $\frac{\pi^2}{15}\,T^4$ (Stefan-Boltzmann law), and 
is bounded from above by $u_{\tiny\mbox{RJ}}$. This provides for an energetic reason why the ``rotation'' from SU(2)$_e$ 
to SU(2)$_{\tiny\mbox{CMB}}$ is invoked in the emergence of blackbody radiation. 
Fixing the critical, thermodynamical temperature $T_c$ for the deconfining-preconfining 
phase transition in SU(2)$_{\tiny\mbox{CMB}}$ as $T_c=T_0=2.725\,$K \cite{lineTemp}, one obtains 
$\frac{1}{\pi}\left(\frac{\Lambda_{\tiny\mbox{CMB}}}{T}\right)^3=1.68\,$GHz. This supports the claim in \cite{lineTemp}
that the CMB radio access, see \cite{Arcade2} and references therein, indeed is attributed to evanescent 
SU(2)$_{\tiny\mbox{CMB}}$ {\sl waves} whose spectral energy density is forced to be maximal at $\nu=0$ 
by a $T$-dependent Meissner mass, signalling the onset of this phase transition. 

To view photons as 
thermal excitations of the thermal ground state in SU(2)$_{\tiny\mbox{CMB}}$ would relate to the photoelectric 
effect in the following way. In SU(2)$_e$ an incident monochromatic wave of frequency, say, $\sim 10^{14}\,$Hz, 
drives the dissipation of radiation-field energy within a thin surface layer of a bulk metal or 
semiconductor (the classical skin effect with skin depth, e.g., in copper, of a few nanometers at $\nu\sim 10^{14}\,$Hz) 
such that an equilibrium between energy entry into this surface layer and heat flow towards the bulk is established. Such an equilibrium 
is characterised by a thermodynamical temperature $T\ll \Lambda_e=m_e$. By the above argument, this 
local thermal environment, however, is subject to 
SU(2)$_{\tiny\mbox{CMB}}$. Since in SU(2)$_{\tiny\mbox{CMB}}$ $s(T)$ in Eq.\,(\ref{sofT}) is much larger 
than $l=\nu^{-1}$ the excitations of the thermal 
ground state are {\sl photons}. That is, the incoming SU(2)$_{e}$ wave of intensity $|\vec{E}_e|^2$ is not supported 
within such a thermal surface layer: by probing the interior 
of SU(2)$_{\tiny\mbox{CMB}}$ calorons/anticalorons
it ``decays'' into {\sl photons} of energy and momentum $2\pi\nu$. With a finite probability \cite{Mandeletal1964}, such a quantum of energy 
and momentum is transferred to a layer electron which, in turn, is expelled from the material thus becoming a photoelectron.  
Modulo a material dependent work function (a function of material parameters such as skin depth, electric and heat conductivity, etc.), the maximal 
kinetic energy of photoelectrons thus is given by $2\pi\nu$ while their flux is proportional to 
the wave's intensity (energy conservation after the above described dynamical equilibrium is established).       

Finally, the thermal ground state of SU(2)$_e$ ($\Lambda=\Lambda_e=m_e$) could explain on a deeper level the transition from 
Thomson (T) scattering of a classical {\sl wave} to Compton (C)
scattering of a {\sl photon} off an electron, which is well described by Quantum Electrodynamics. The associated total cross section \cite{KleinNishima} -- an intensity independent quantity -- is given as
\eab
\label{compton}
\sigma_C&=&\frac34\sigma_T\left[\frac{1+x}{x^3}\left(\frac{2x(1+x)}{1+2x}-\log(1+2x)\right)+\frac{1}{2x}\log(1+2x)-\frac{1+3x}{(1+2x)^2}\right]\nonumber\\ 
&=&\sigma_T[1-2x+\frac{26}{5}x^2+O(x^3)]\,,\ \ \ (x\equiv\nu/m_e)\,.
\eae
As $x$ rises to order unity, an increasingly strong violation of (\ref{uncert}) 
takes place \footnote{Since $\sigma_C$ is defined as an intensity independent quantity the level of 
violation of (\ref{uncert}) is to be inferred at {\sl constant}  
$|\vec{E}_e|^4$. In a nonthermal situation $\Lambda_e=m_e$ is the only mass scale in SU(2)$_e$, and 
thus it is natural to set $|\vec{E}_e|^4=m_e^8$. As a consequence, (\ref{uncert}) requires $\nu\ll 8\,m_e$ to obtain a beam 
void of particle-like aspects.}. 
Here the scattering off an isolated electron does not generate any local, thermal equilibrium 
but the incoming beam itself is of an increasingly corpuscular structure as $x$ grows. 

To summarise, we have discussed how condition (\ref{uncert}) for the wave-like propagation of 
electromagnetic disturbances, demanding that caloron/anticaloron centers in the thermal ground state
are not excited \cite{GrandouHofmann2015}, and the violation of this condition may underly wave-particle 
duality of electromagnetic disturbances in their propagation and interaction with the stable electric charge of the electron. 
A single SU(2) Yang-Mills theory, conjectured to describe the thermodynamical situation -- SU(2)$_{\tiny\mbox{CMB}}$ \cite{Hofmann2005,lineTemp} --,   
does not in general account for wave-like, nonthermal propagation. Therefore, a product of (at least) two mixing SU(2) Yang-Mills 
theories of hierarchically different Yang-Mills scales had to be postulated. Our brief conceptual discussion 
necessarily leaves a number of important questions unanswered. 
For example, is the liberation of a quantum of energy $2\pi\nu$ within a caloron/anticaloron 
center by an external drive field of frequency $\nu$ understandable 
as a resonant excitation involving monopole shaking \cite{GPY1982,GrandouHofmann2015}, 
which is resolved by this very liberation? Or, how can one accomodate the electroweak interactions between neutral and unstable 
charged leptons -- very successfully and effectively 
described by the present Standard Model of Particle Physics -- in a framework of 
mixing, {\sl pure} SU(2) Yang-Mills theories. And how would this extend to incorporate the electroweak 
interactions of hadrons. We hope to be able to shed more light on such questions in the future.

\end{document}